\pdfoutput=1
\documentclass[10pt, letterpaper]{article}
\usepackage[textheight=9in,textwidth=6.5in,top=.9in]{geometry}
\usepackage{framed,multirow}
\usepackage{amssymb}

\usepackage{url}

\usepackage{hyperref}
\usepackage{siunitx}
\usepackage{tabularx}
\usepackage{threeparttable}
\usepackage{graphicx}
\newcolumntype{L}{>{\raggedright\arraybackslash}X}
\newcolumntype{R}{>{\raggedleft\arraybackslash}X}
\newcolumntype{C}{>{\centering\arraybackslash}X}
\usepackage{booktabs}
\usepackage[utf8]{inputenc}
\DeclareUnicodeCharacter{2212}{-}
\usepackage{lmodern}
\usepackage{chngcntr}
\usepackage{natbib}

\title{Bayesian Pharmacokinetic Modeling of Dynamic Contrast-Enhanced Magnetic Resonance Imaging: Validation and Application}

\author{Andreas Mittermeier \thanks{Department of Radiology, Ludwig-Maximilians-University Hospital Munich, 81377 Munich, Germany} \and
Birgit Ertl-Wagner\footnotemark[1] \thanks{Department of Neuroradiology, The Hospital for Sick Children, Toronto ON M5G 1X8, Canada} \and
Jens Ricke\footnotemark[1] \and
Olaf Dietrich\footnotemark[1] \and
Michael Ingrisch\footnotemark[1]}

\begin{document}
	
\date{}
\maketitle

{\bfseries \small \textit{Abstract} - Tracer-kinetic analysis of dynamic contrast-enhanced magnetic resonance imaging data is commonly performed with the well-known Tofts model and nonlinear least squares (NLLS) regression. This approach yields point estimates of model parameters, uncertainty of these estimates can be assessed e.g. by an additional bootstrapping analysis. Here, we present a Bayesian probabilistic modeling approach for tracer-kinetic analysis with a Tofts model, which yields posterior probability distributions of perfusion parameters and therefore promises a robust and information-enriched alternative based on a framework of probability distributions.
In this manuscript, we use the  Quantitative Imaging Biomarkers Alliance (QIBA) Tofts phantom to evaluate the Bayesian Tofts Model (BTM) against a bootstrapped NLLS approach. Furthermore, we demonstrate how Bayesian posterior probability distributions can be employed to assess treatment response in a breast cancer DCE-MRI dataset using Cohen’s \textit{d}.
Accuracy and precision of the BTM posterior distributions were validated and found to be in good agreement with the NLLS approaches, and assessment of therapy response  with respect to uncertainty in parameter estimates was found to be excellent.
In conclusion, the Bayesian modeling approach provides an elegant means to determine uncertainty via posterior distributions within a single step and provides honest information about changes in parameter estimates.

}

\section{Introduction}
Dynamic contrast-enhanced magnetic resonance imaging (DCE-MRI) is a noninvasive imaging technique used to quantify microvascular tissue perfusion with the help of a contrast agent (CA) \citep{ingrisch_tracer-kinetic_2013}. In MRI, a gadolinium-based CA is used most commonly and injected intravenously after the acquisition of pre-contrast baseline scans. The CA increases T1 and T2 relaxation rates of surrounding water protons and causes signal enhancement in a T1-weighted acquisition. By measuring multiple $T_1$-weighted images during the passage of the CA through the tissue of interest, a time-dependent CA concentration can be extracted from the signal-time course of each voxel. Besides determining semi-quantitative and descriptive parameters from the concentration curves, e.g. time to peak, area under curve, or maximum, quantitative perfusion parameters can be obtained by fitting pharmacokinetic (PK) models to the data \citep{sourbron_tracer_2012, sourbron_classic_2013, roberts_comparative_2006}. Popular PK models that characterize CA transport from DCE-MRI data are the classical Tofts model (TM) \citep{tofts_modeling_1997}, the extended Tofts model and the two compartment exchange model \citep{sourbron_scope_2011}.

The standard approach for estimating PK parameters from DCE-MRI data is using non-linear regression to determine a maximum likelihood estimator by non-linear least squares (NLLS) analysis \citep{seber_nonlinear_2003}. For this purpose, an optimizing algorithm aims to minimize the sum of squared residuals between model and data and yields, if successful, a \textit{point estimate} of model parameters. The NLLS approach is widely used, and a number of software packages provide non-linear regression implementation of a range of PK models \citep{huang_variations_2014, beuzit_dynamic_2016}. Bayesian probabilistic modeling, on the other hand, offers an alternative modeling approach within a framework of probability distributions. Briefly, a \textit{prior} belief about model parameters is formulated as a probability distribution; this allows to incorporate domain expertise, e.g. physical constraints. With dedicated algorithms, this prior belief is then updated with the measured data and yields the \textit{posterior} probability distributions of the parameters given the data \citep{mcelreath_statistical_2015}. Through recent algorithmic developments \citep{hoffman_no-u-turn_2011} and the increasing availability of computational power, the use of Bayesian modeling approaches is spreading in various disciplines and has already shown to be a robust and accurate alternative for the analysis of MR imaging data \citep{schmid_bayesian_2006,orton_bayesian_2007,woolrich_bayesian_2009,dikaios_comparison_2017,tietze_bayesian_2018,hansen_robust_2019}. The posterior probability distributions that result from Bayesian modeling greatly increase the interpretability of analysis results. Compared to simple point estimates, entire parameter probability distributions allow a straightforward assessment of, e.g.,  whether a parameter has truly changed in the course of a therapy, or whether the parameter change has only occurred within the uncertainty of the estimation \citep{shukla-dave_quantitative_2018}.

In the present manuscript, we investigated Bayesian tracer-kinetic modeling in the context of DCE-MRI. To this end, we implemented a Bayesian TM (BTM) with the purpose to i) evaluate accuracy against a NLLS approach using a digital reference object, ii) validate uncertainty estimates against a bootstrapped NLLS approach to assess the precision and iii) demonstrate how Bayesian posterior probability distributions can be used to assess treatment response in a breast cancer DCE-MRI dataset.


\section{Materials and Methods}
\subsection{Signal Conversion and Pharmacokinetic Models}
In a typical DCE-MRI experiment, time-resolved signal intensity curves $S(t)$ are extracted voxel-wise from multiple $T_1$-weighted images. To derive quantitative information, the measured signal intensities need to be converted to CA concentration curves. For this purpose, the signal equation for the spoiled gradient echo (SPGR) sequence in steady state can be used with the baseline signal $S_0(t)$, flip angle $\alpha$, repetition time $T_R$ and relaxation rate $R_1(t)$ as:
\begin{equation}
S(t)=S_0\sin(\alpha)\frac{1-e^{-T_RR_1(t)}}{1-\cos(\alpha) e^{-T_RR_1(t)}} .
\label{eq:spgr}
\end{equation}
One can solve Eq. (\ref{eq:spgr}) for the time-dependent relaxation rate $R_1(t)$:
\begin{equation}
R_1(t) = -\frac{1}{T_R} \log \left( \frac{1-A}{1-\cos(\alpha)A} \right),
\label{eq:R1}
\end{equation}
with the auxiliary variable 
\begin{equation}
A=\frac{S(t)}{S_0}\frac{1-e^{-T_R R_{10}}}{1-cos(\alpha)e^{-T_R R_{10}}}.
\end{equation}
A time-dependent concentration can then be calculated from the linear relation to the change in relaxation rates during and before administration of CA, $R_1(t)$ and $R_{10}$, respectively:
\begin{equation}
c(t)=(R_1(t)-R_{10})/r_1,
\label{eq:change_R}
\end{equation}
with the specific \textit{relaxivity} of the gadolinium-based CA $r_1$ \citep{pintaske_relaxivity_2006}.

A standard approach for the analysis of concentration-time curves in DCE-MRI data is the TM \citep{tofts_measurement_1991,tofts_modeling_1997,sourbron_scope_2011} which assumes a negligible amount of intravascular tracer and describes CA transportation as:
\begin{equation}
c_t(t) = K^\text{trans} e^{-tk_\text{ep}} * c_p(t).
\label{eq:toftsmodel}
\end{equation}
Here, $c_t(t)$ is the time-dependent concentrations of CA in the tissue of interest; $c_p(t)$ is the concentration in the blood plasma of the tissue-feeding artery, often referred to as arterial input function (AIF). $c_t(t)$ and $c_p(t)$ are connected with a convolution, expressed as ``$*$''. The parameter $v_e$ is the volume fraction of the interstitium, the extravascular extracellular space (EES). $K^\text{trans}$ is defined as the transfer constant of CA between blood plasma and EES. The rate constant $k_\text{ep}=K^\text{trans}/v_e$ is the ratio of the transfer constant to the EES \citep{tofts_estimating_1999,sourbron_scope_2011}.

The tissue concentration $c_t(t)$ can be calculated from the measured signal $S(t)$ with Eq. (\ref{eq:spgr}-\ref{eq:change_R}) using the relaxation time $T_{10}$ in tissue via $R_{10} = 1 / T_{10}$. Plasma concentration $c_p(t)$ in the AIF can be calculated likewise using the relaxation time $T_{10}$ of blood and the additional transformation from blood to plasma concentration via the hematocrit $hct$:
\begin{equation}
c_p(t) = c_b(t) \cdot (1-hct).
\label{eq:plasma_conc}
\end{equation}

The standard TM is used in the following within a classical NLLS likelihood framework and a Bayesian framework to quantify perfusion in simulated and measured DCE-MRI data. To account for noise in any observed data, an error term is added to the TM from Eq. (\ref{eq:toftsmodel}) and an observation $i$ is given as
\begin{equation}\label{eq:tm_stat_model_errorterm}
y_i = c_t(t_i, \theta)+\epsilon_i .
\end{equation}
where the model parameters $K^\text{trans}$ and $v_e$ are summarized in the vector $\theta$.

\subsection{Data}
\subsubsection{Validation: QIBA DCE-MRI Phantom}
\begin{figure*}[ht!]
	\includegraphics[width=\textwidth]{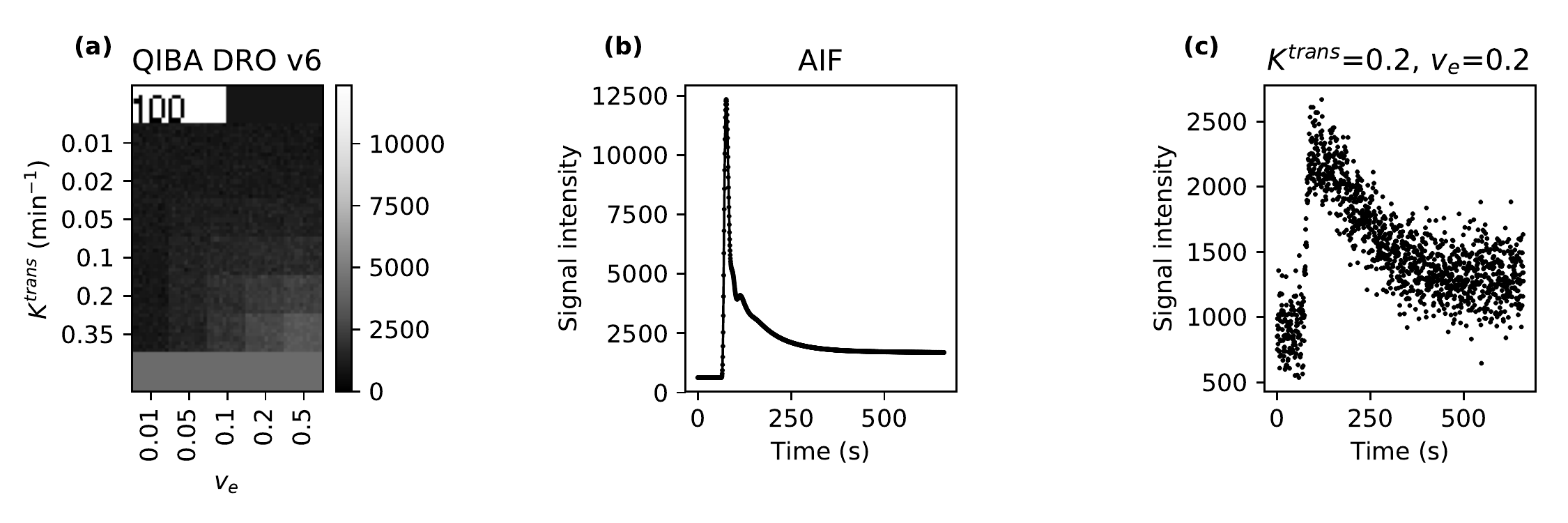}
	\caption{\textbf{(a)} Snapshot of the \textit{QIBA\_v6\_Tofts} DRO at $t=100$ s; the AIF is the bottom strip of the image, maximum of the AIF with timepoint labels in seconds are the top left strip and the zero patch ($K^\text{trans}=0.0$ min$^{-1}$, $v_e=0.5$) is the top right strip. Intensity-time curves for AIF \textbf{(b)} and one pixel with $K^\text{trans}$=0.2 min$^{-1}$ and $v_e$=0.2 with added noise \textbf{(c)}, respectively.}
	\label{fig:DROv6}
\end{figure*}

To evaluate accuracy of estimates and compare results of different fitting approaches, a simulated phantom with known PK parameters was investigated first. The Quantitative Imaging Biomarkers Alliance (QIBA)\footnote{\url{https://sites.duke.edu/dblab/qibacontent/}} provides several freely available test images for DCE-MRI analysis, known as digital reference objects (DRO). These have been used previously to validate various fitting algorithms and analysis toolkits \citep{ortuno_dceurlab:_2013,smith_dcemri.jl:_2015,debus_mitk-modelfit:_2019}. The noise-free \textit{QIBA\_v6\_Tofts} version was chosen here. The DRO contains simulated DCE-MRI data generated with the standard TM in Eq. (\ref{eq:toftsmodel}) for a study duration of $t=660$ s with a temporal resolution $\Delta t=0.5$ s. Tissue concentration-time curves $c_t(t)$ have been created for all combinations of $K^\text{trans} \in$ \{0.01,0.02,0.05,0.1,0.2,0.35\} min$^{-1}$ and $v_e \in$  \{0.01,0.05,0.1,0.2,0.5\}, filling a 10$\times$10 pixel patch for each combination. Table \ref{tab:parameters} lists the parameters stated in the QIBA description\footnote{\url{https://sites.duke.edu/dblab/files/2015/05/Dynamic\_v6\_beta1\_description\_Rev1.pdf}}, following QIBA's DCE MRI quantification profile\footnote{\url{http://qibawiki.rsna.org/images/1/12/DCE-MRI\_Quantification\_Profile\_v1.0.pdf}} to convert signal intensities to concentrations (compare Eq. (\ref{eq:spgr}-\ref{eq:change_R})). For a more realistic setting, complex Gaussian noise with standard deviation $\sigma=0.2$ relative to the pre-contrast baseline signal $S_0$ was added to the original noise-free test data. No noise was added to the AIF for simplicity and to be able to reliably relate our results to published work of \citet{smith_dcemri.jl:_2015} and \citet{ortuno_dceurlab:_2013}. Fig. \ref{fig:DROv6} shows a snapshot of the DRO  signal intensities at $t=100s$, the AIF and an exemplary voxel with parameters $K^\text{trans}=0.2$ min$^{-1}$ and $v_e=0.2$, respectively.

\subsubsection{Application: Breast Cancer DCE-MRI Data}
The Quantitative Imaging Network (QIN) aims at improving quantitative imaging and does so by sharing data which was acquired as part of various QIN studies, collected in The Cancer Imaging Archive (TCIA)\footnote{\url{https://wiki.cancerimagingarchive.net/display/Public/Collections}} \citep{clark_cancer_2013}. A set of breast cancer DCE-MRI data \citep{huang_variations_2014-1} in DICOM format acquired from 10  patients was used to demonstrate the performance of the BTM on clinical data. The dataset contains DCE-MRI measurements  acquired before  (visit 1) and during (visit 2) preoperative neoadjuvant chemotherapy (NACT), respectively. For three patients,  pathologic complete response (pCR) was reported, the remaining seven patients were classified as non-pCR. In addition, the dataset includes a region of interest (ROI) per patient, drawn by an experienced breast radiologist. A sample-averaged AIF is provided as blood concentration $c_b(t)$ and was converted to plasma concentration $c_p(t)$ using Eq. (\ref{eq:plasma_conc}). Signal intensities within the ROI were converted to tissue concentrations using Eq. (\ref{eq:spgr}-\ref{eq:change_R}). Parameters for the conversion are specified in Table \ref{tab:parameters}, further details can be found in the original work by \citet{huang_variations_2014}.

\begin{table*}[ht!]
	\centering
	\begin{threeparttable}
		\caption{Parameters for the conversion from signal to concentration}
		\label{tab:parameters}
		\begin{tabular}{lccccccc}
			\toprule
			& {${T_{10}}$(Tissue)} & {${T_{10}}$(Blood)} & {${r_1}$} & {${\alpha}$} & {${T_R}$}  & {${T_E}$} &{${hct}$} \\
			\midrule 
			QIBA DRO\tnote{1}   & 1000 ms          & 1440 ms         & 0.0045 Lmmol$^{-1}$ms$^{-1}$ & 30$^{\circ}$ & 5 ms   & -      & 0.45  \\ 
			QIN Breast\tnote{2} & 1666 ms\tnote{3}          & 1440 ms         & 0.0045 Lmmol$^{-1}$ms$^{-1}$ & 10$^{\circ}$ & 6.2 ms & 2.9 ms & 0.45  \\ 
			\bottomrule
		\end{tabular}
		\begin{tablenotes}\footnotesize
			\item[1] Quantitative Imaging Biomarker Alliance Digital Reference Object $QIBA\_v6\_Tofts$
			\item[2] Quantitative Imaging Network Breast Cancer Dataset
			\item[3] Personal communication with the author of \citet{huang_variations_2014}
		\end{tablenotes}
	\end{threeparttable}
\end{table*}

\subsection{Models and Analysis}
\subsubsection{Non-linear Least Squares Approach with Bootstrapping}
The standard evaluation of DCE-MRI data is performed in a likelihood framework by fitting a non-linear regression model to the concentration-time curve in every voxel. The NLLS approach minimizes the sum of squared errors between measured data $y_i$ at timepoint $t_i$ for $i=0,1,...,N$ and the model function $c_t(t_i)$ in Eq. (\ref{eq:tm_stat_model_errorterm})
\begin{equation}
\min\sum_{i=0}^N(y_i-c_t(t_i, \theta))^2=\min\sum_{i=0}^N\epsilon_i^2
\label{eq:min_sse}
\end{equation}
to infer the \textit{best guess} parameter $\hat{\theta}$. Assuming normally distributed noise $\epsilon_i$, the least-squares estimator $\hat{\theta}$ equals the maximum-likelihood estimator \citep{seber_nonlinear_2003}. 

An implementation of the Broyden-Fletcher-Goldberg-Shano (L-BFGS) algorithm \citep{byrd_limited-memory_1994,zhu_algorithm_1997} in \textit{SciPy}\footnote{Python 3.6.6, scipy 1.1.0, \url{https://www.scipy.org/}} \citep{jones_scipy:_2001} was used for inference of the parameters via the \textit{optimize.minimize} function. Initial values for $K^\text{trans}$ and $v_e$ were set to 0.001. The concentration curves of the DRO were then fitted and parameter maps were constructed for $K^\text{trans}$ and $v_e$. By comparing them to the true parameter maps, percentage error maps were calculated as $\theta_\text{\%err}=(\hat{\theta}-\theta_\text{true})/\theta_\text{true}$.

A bootstrap method was implemented to assess the uncertainty of $\hat{\theta}$ \citep{kershaw_precision_2006}. For that, the residuals, i.e. the difference between the fitted and the measured curve were calculated. In a next step, the residuals were resampled by randomly drawing samples with replacement. Subsequently, the resampled residuals were added to the fitted curve and the TM was used to determine another set of estimates, equivalent to inferring the original best guess. The number of iterations was set to 1000.

Uncertainty maps were then calculated from the bootstrap samples for the NLLS approach. Denoted as $\sigma$, half the width between 17th and 83rd percentile was considered a more robust measure for the precision than the standard deviation and is used throughout this work. For samples following a Gaussian normal distribution, $\sigma$ would be equal to the standard deviation.

\subsubsection{Bayesian Inference and Implementation}
The alternative evaluation is performed in a Bayesian framework which infers a full posterior distribution $P(\theta\mid y)$ of the model parameters $\theta$ given an observation of data $y$. The observational error $\epsilon_i$ for each measurement $y_i$ at timepoint $t_i$ for $i=0,1,...,N$ in Eq. (\ref{eq:tm_stat_model_errorterm}) is assumed to be Gaussian with standard deviation $\sigma$. Hence, the joint observations of CA concentration in each voxel, conditional on the parameters, are modeled in the \textit{likelihood} as
\begin{equation}
P(y\mid\theta) = \prod_{i=0}^N  \mathcal{N}(y_i \mid c_t(t_i, \theta), \sigma^2),
\end{equation}
with $\mathcal{N}$ representing a normal distribution and $c_t(t_i,\theta)$ the CA tissue concentration evaluated with the TM in Eq. (\ref{eq:toftsmodel}).

Information about the parameters prior to the observation of data are specified in the \textit{prior distribution} $P(\theta)$, enforcing physical or biological constraints. The likelihood of the data $P(y\mid\theta)$ and the product of the prior probability densities $P(\theta)$ are combined with the observed data to infer the joint posterior distribution via Bayes' theorem:
\begin{equation}\label{eq:BayesTheorem}
P(\theta\mid y)=\frac{P(y\mid \theta)\,P(\theta)}{P(y)}.
\end{equation}

The denominator in Eq. (\ref{eq:BayesTheorem}) is referred to as model evidence and calculates as $P(y)=\int P(\theta)P(y\mid\theta)\mathrm{d}\theta$. If the complexity of the model allows no analytical solution to this integral, Markov Chain Monte Carlo (MCMC) methods \citep{gilks_markov_1995} offer a means to determine the posterior probability distribution. Briefly, a MCMC algorithm draws samples from a target distribution, which equals the desired posterior distribution. The accepted parameter proposals are stored in a chain or trace of estimates \citep{kruschke_doing_2014}.

The BTM was implemented in \textit{Stan} \citep{carpenter_stan:_2017}, an open-source software package, using \textit{pystan}\footnote{Python 3.6.6, pystan 2.18.0, \url{https://pystan.readthedocs.io/}}. In the present analysis, weakly informative priors were chosen. In particular, for the volume fraction $v_e \in [0,1]$ a beta prior $v_e \sim \mathrm{Beta}(\alpha=2,\beta=2)$ was chosen and for $K^\text{trans} \in \mathbb{R}_+$ a gamma prior was specified $K^\text{trans} (\text{min}^{-1}) \sim \mathrm{Gamma}(\alpha=1.1, \beta=1/0.002)$. The prior for the standard deviation of the observational error was set to $\sigma (\text{mmol/L}) \sim \mathrm{LogNormal}(\mu=0, \sigma=1)$. The appendix \ref{app:prior_p_c} provides a prior predictive check on these prior distributions. MCMC samples were drawn from the posterior distribution with the No-U-Turn (NUTS) algorithm \citep{hoffman_no-u-turn_2011}. The number of iterations was set to 1000, sampled in two chains simultaneously, following a warm-up period of 500 iterations. Stan also reports divergences of the sampling algorithm and indicates the need to update the default settings of NUTS, e.g. initial step size and target acceptance rate.

To monitor the convergence of the MCMC chains to the target distribution, different diagnostics are automatically run alongside in Stan. The potential scale reduction statistic, $\hat{R}$, by \citet{gelman_inference_1992} compares the sample variance within and across chains, and indicates if chains have not converged to a common distribution ($\hat{R}>1.1$). The effective sample size $N_{\text{eff}}$ indicates the degree of uncertainty in estimates due to autocorrelation of samples \citep{geyer_introduction_2011}.

All concentration curves of the DRO were then fitted with the BTM to obtain posterior probability distributions of the parameters $\theta$. To be able to compare the distributions to point estimates and to generate parameter maps, two hallmarks of the posterior distributions were determined: the median and, as for the bootstrap samples, half the distance between the 17th and 83rd percentile, denoted as $\sigma$. By comparing the median parameter maps to the true parameter maps, a map of the percentage error  was calculated as above to assess the accuracy of estimates.

To evaluate the breast cancer DCE-MRI datasets, the mean tissue concentration curve over the ROI $c_{t,\text{ROI}}(t)$ was calculated for each patient and both visits. Subsequently, all concentration-time curves were fitted with the BTM to infer posterior distributions for the model parameters $\theta$. To ensure that the model adequately captured the underlying data generating process, a \textit{posterior predictive check} (PPC) was performed. Briefly, we used the BTM to generate new predictive data $\hat{y}$ and checked if it resembled the observed data. The full posterior distribution is exploited in this way to generate a \textit{posterior predictive distribution}
\begin{equation}
P(\hat{y}|y)=\int P(\hat{y}|\theta)P(\theta|y)\mathrm{d}\theta,
\label{eq:ppc}
\end{equation}
which propagates the uncertainty in the parameter estimates to uncertainty about prediction \citep{betancourt_unified_2015,mcelreath_statistical_2015,gabry_visualization_2017}. In this way, PPCs allow to detect systematic modeling errors and violations of  model assumptions. Subsequently, the posterior distributions of $K^\text{trans}$ were compared across visits for all patients with the objective to discriminate between patients with pCR and non-pCR.

\subsection{Statistical Analysis}
A quantitative statistical measure for signal fidelity is the structural similarity index (SSIM) \citep{wang_image_2004}. It gives an average value over similarities of three key elements of an image: luminance, contrast and structure \citep{zhou_wang_mean_2009}. To assess the accuracy of parameter estimates for the DRO, the SSIM was calculated between the estimated and the true parameter maps. As a comparison, the root-mean-squared error (RMSE) was calculated alongside. In order to get reasonable values for RMSE, outliers in $K^\text{trans}$-estimates obtained from NLLS fitting needed to be restricted to one. In addition, the SSIM was calculated between the $\sigma$-uncertainty maps determined with the BTM and the bootstrapping method to assess similarities in the precision of estimates.

To compare the $K^\text{trans}$ posterior distributions between visits for the breast cancer DCE-MRI dataset, Cohen's $d$ was calculated for each of the ten patients as:
\begin{equation}
d={\frac  {{\bar{x}}_{1}-{\bar{x}}_{2}}{{\sqrt{(\sigma_{1}^{2}+\sigma_{2}^{2})/2}}}}.
\end{equation}
$\bar{x}$ represents the average $K^\text{trans}$ value per visit, $\sigma$ its standard deviation. In this way, the width of the posterior distributions are incorporated into a single value. Compared to just reporting the percentage change of $K^\text{trans}$ mean values, the uncertainty in parameter estimation is accounted for. An univariate logistic regression (ULR) model, implemented in \textit{scikit-learn}\footnote{Python 3.6.6, scikit-learn 0.20.0, \url{https://scikit-learn.org/}} \citep{pedregosa_scikit-learn:_2011}, was fitted to the Cohen's $d$ values. The receiver operating characteristic (ROC) area under curve (AUC) was calculated in order to obtain a quantitative measure for the assessment of response.


\section{Results}
\subsection{Validation: QIBA DCE-MRI Phantom}
\begin{figure*}[ht!]
	\includegraphics[width=1.\linewidth]{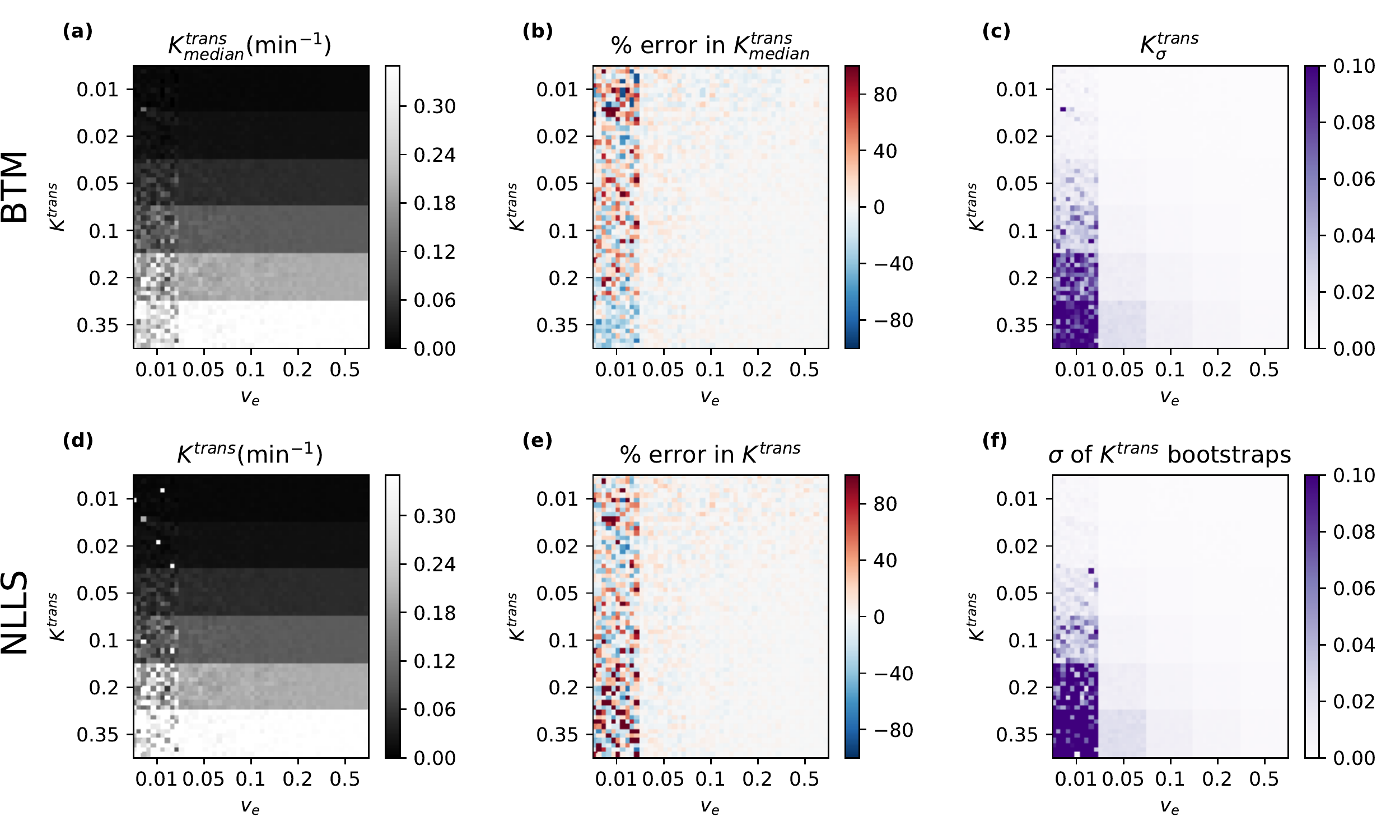}
	\caption{Estimated $K^\text{trans}$ with BTM (top) and NLLS approach (bottom). The left column displays median \textbf{(a)} and point estimates \textbf{(d)}. The middle column \textbf{(b,e)} shows the calculated percentage error between estimates and ground truth. The right column illustrates the uncertainty $\sigma$ of the Bayesian posterior \textbf{(c)} and the additional bootstrap samples \textbf{(f)}.}
	\label{fig:parameter_maps}
\end{figure*}

Concentration-time curves of the DRO were evaluated within a Bayesian and likelihood framework. The resulting parameter estimates for $K^\text{trans}$ are exemplarily shown in Fig. \ref{fig:parameter_maps} for the Bayesian approach \textbf{(a)} and the NLLS reference \textbf{(d)}. Note that the voxels in the Bayesian framework show median values of their respective posterior distributions while voxels in the likelihood framework represent point estimates. In general, the parameter maps show high accordance with the true values. The corresponding percentage error maps in the middle column \textbf{(b)} and \textbf{(e)} display relatively low errors for all regions with $v_e > 0.01$ for both methods. Low accuracy, hence high percentage errors are observed for regions where $v_e = 0.01$. SSIM between estimated and true $K^\text{trans}$-maps is higher for the BTM than for the NLLS approach. Furthermore, RMSE is lower for the BTM for both PK parameter maps. Details are provided in Table \ref{tab:stat_results}.

\begin{table*}[ht!]
	\centering
	\begin{threeparttable}
	\caption{SSIM and RMSE between estimated DRO $K^{trans}$ and $v_e$ parameter maps and ground truth for both approaches; SSIM of 100\% indicates perfect similarity.}
	\label{tab:stat_results}
	\begin{tabularx}{0.7\textwidth}{lCCCC}
		\toprule
		& \multicolumn{2}{c}{$K^\text{trans}$} & \multicolumn{2}{c}{$v_e$} \\
		& {BTM} & {NLLS} & {BTM} & {NLLS}       \\
		\midrule
		SSIM & 96 \%              & 91 \%               & 92 \%           & 94 \%            \\
		RMSE & 2.5 \%             & 7.0 \%              & 4.1 \%          & 5.4 \%           \\
		
		\bottomrule
	\end{tabularx}
	\begin{tablenotes}\footnotesize
		\item[] BTM = Bayesian Tofts model; NLLS = Non-linear least squares approach; SSIM = Structural similarity index; RMSE = Root-mean-squared error
	\end{tablenotes}
	\end{threeparttable}
\end{table*}

The right column of Fig. \ref{fig:parameter_maps} displays the precision of the parameter estimates evaluated with the BTM \textbf{(c)} and a bootstrapping method applied to the fit results of the NLLS approach \textbf{(f)}. The visual analysis of the uncertainty maps reveals very similar patterns for both approaches, supported by a SSIM of 91\%. The highest uncertainty  occurs in regions with the highest percentage error for the fitting parameter estimates. The remaining parameter combinations have much greater precision. Information about divergences (BTM) and pixels where the NLLS algorithm did not find a solution can be found in Table \ref{tab:fiterrors}, together with computational times for fitting all 3000 pixels with BTM and NLLS approaches and the additional bootstrap analysis.

\begin{table*}[ht!]
	\centering
	\begin{threeparttable}
		\caption{Fitting process and parameter estimation of all 50$\times$60 DRO concentration curves}
		\label{tab:fiterrors}
		\begin{tabularx}{0.7\textwidth}{lCC}
			\toprule
			& {BTM}              & {NLLS}                 \\
			\midrule
			Divergences  & 17               & 27                    \\
			Computational time: fitting & $\sim 48$ min &  $\sim 2$ min \\
			Computational time: uncertainty & \textit{included} & $\sim 2100$ min\tnote{*} \\
			\bottomrule
		\end{tabularx}
		\begin{tablenotes}\footnotesize
			\item BTM = Bayesian Tofts Model; NLLS = Non-linear Least Squares approach
			\item[*] Based on additional bootstrap analysis
		\end{tablenotes}
	\end{threeparttable}
\end{table*}

\subsection{Application: Breast Cancer DCE-MRI Data}
\begin{figure*}[ht!]
	\includegraphics[width=\linewidth]{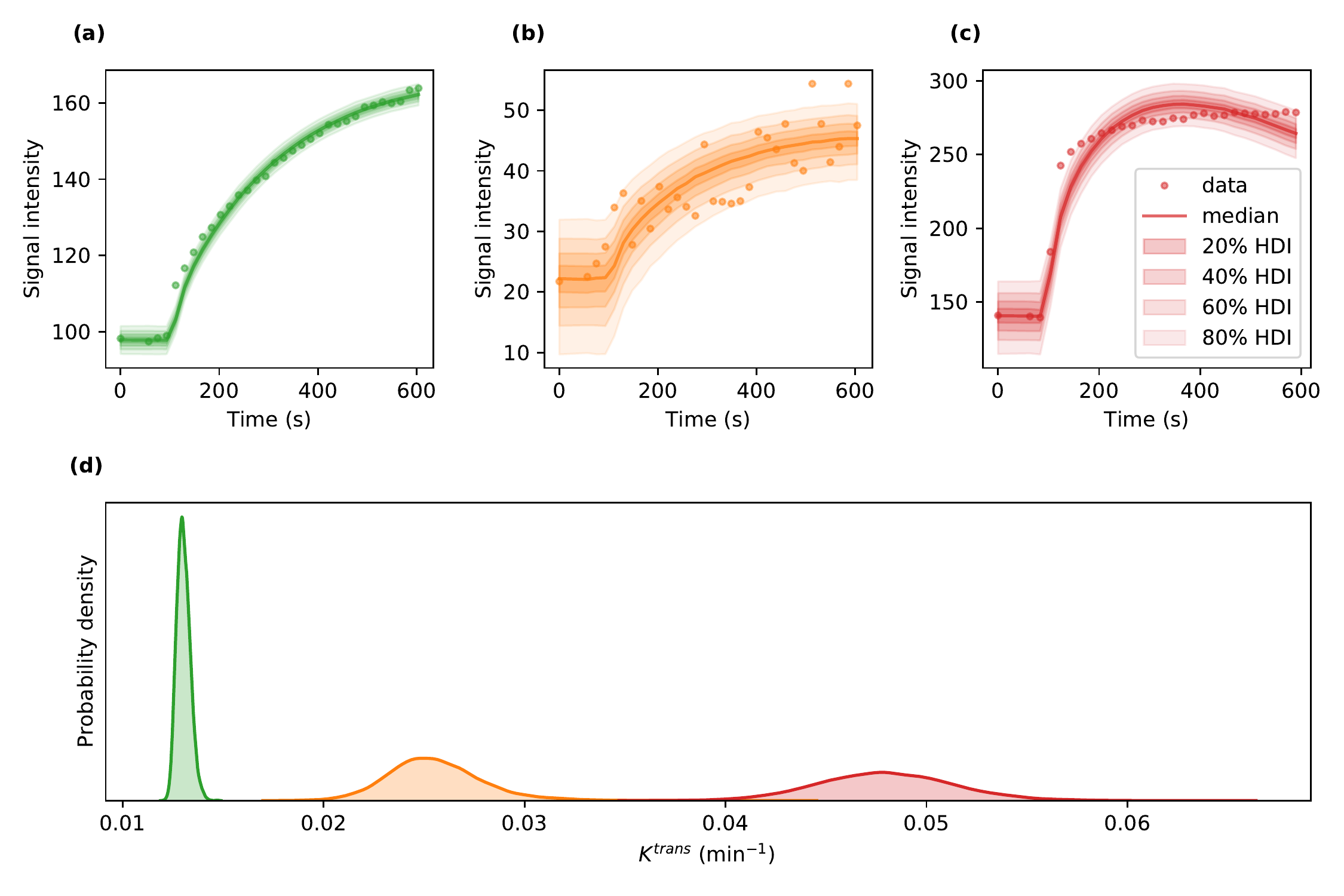}
	\caption{Observed signal intensity-time courses (dots) with posterior predictive distribution median (line) and the 20\%, 40\%, 60\% and 80\% highest density intervals (HDI); \textbf{(a)} for a good fit to data with low noise level, \textbf{(b)} for a good fit to data with high noise level, and \textbf{(c)} for a bad fit to data with low noise level. \textbf{(d)} Posterior distributions for $K^\text{trans}$ estimated from the respective signal intensity-time curves (color-coded).}
	\label{fig:3curves}
\end{figure*}

Fig. \ref{fig:3curves} shows  representative signal intensity-time curves with the associated PPCs \textbf{(a-c)} and their corresponding $K^\text{trans}$ posterior distributions \textbf{(d)}. Here, the dark line illustrates the median and the increasingly lighter bands are the 20\%, 40\%, 60\% and 80\% highest density intervals (HDI) between the corresponding (0.4,0.6), (0.3,0.7), (0.2,0.8) and (0.1,0.9) percentiles of the posterior predictive distribution. The PPC in \textbf{(a)} indicates a good fit of the model to the data, the corresponding posterior distribution (green) for $K^\text{trans}$ is narrow. The PPC in \textbf{(b)} suggests that the chosen model provides a good fit to the data, the high noise level in the data is associated with a broader posterior distribution (orange). In \textbf{(c)}, the noise level of the data is comparable to \textbf{(a)}, however the PPC indicates a modeling error.

\begin{figure*}[ht!]
	\includegraphics[width=\linewidth, trim={0 0 0 2cm},clip]{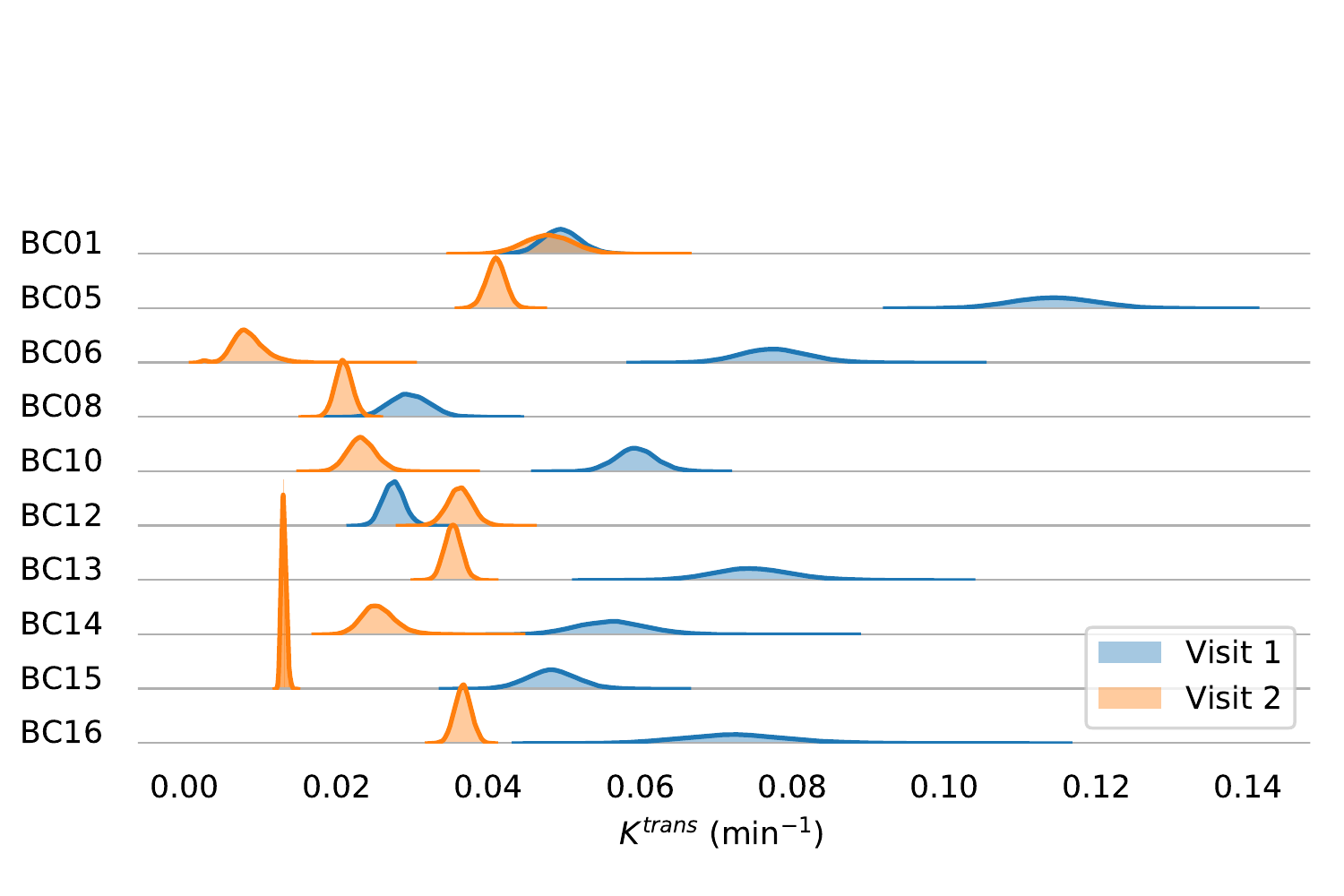}
	\caption{Posterior probability densities of $K^\text{trans}$ for all patients. Blue corresponds to visit 1, orange to visit 2. BC05, BC06 and BC15 are labeled pCR, the rest non-pCR.}
	\label{fig:curves_dists}
\end{figure*}

\begin{figure*}[ht!]
	\includegraphics[width=\linewidth]{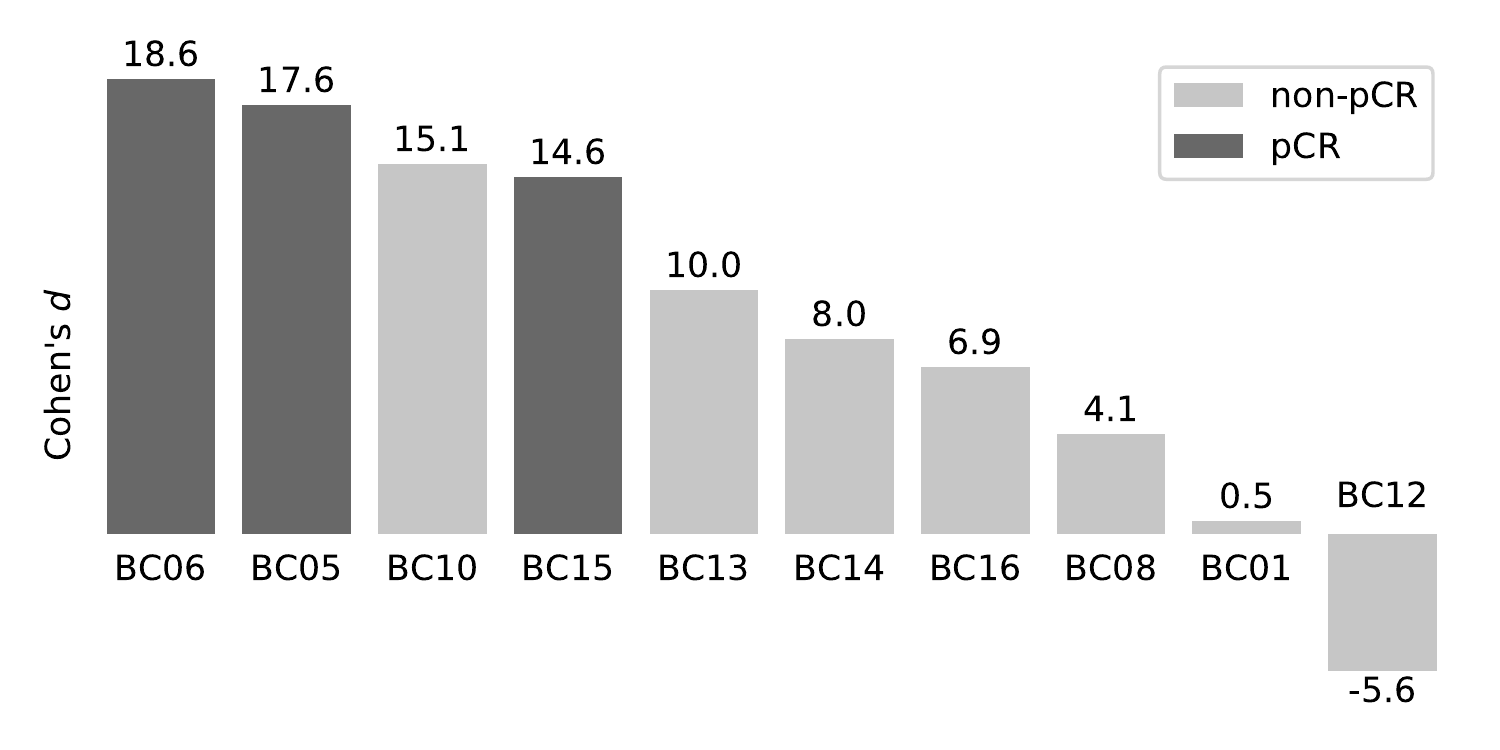}
	\caption{Cohen's $d$ calculated from $K^\text{trans}$ posterior distributions for all patients; sorted by value.}
	\label{fig:cohens_d}
\end{figure*}

Fig. \ref{fig:curves_dists} shows the posterior distributions of $K^\text{trans}$ for all patients for visit 1 (blue) and visit 2 (orange), before and during NACT, respectively. With one exception, a general decrease in $K^\text{trans}$ is observed. The degree of change, dependent on the width of the posterior distributions, is summarized in Cohen’s $d$ values and visualized in Fig. \ref{fig:cohens_d}; light-gray represents non-pCR, dark-gray pCR. The ULR analysis revealed a ROC AUC of 0.952. Computational time for fitting all 20 ROI-averaged concentration curves was $\sim 20$ s for the BTM.

\section{Discussion}
In this study, we assessed posterior probability distributions of tracer-kinetic parameters obtained with a BTM against a standard NLLS approach. Validation with a DRO revealed high accuracy of BTM and NLLS approaches, indicated by strong similarity between estimated and ground truth maps. In addition, precision of estimates, assessed via the width of the posterior probability distributions and bootstrapping, respectively, was in very good agreement between both approaches. Analysis of the breast cancer DCE-MRI dataset with the BTM revealed that the degree of decrease in $K^\text{trans}$ gives information about the pathologic response to NACT. The response in dependence of the uncertainty of parameter estimates was quantified with Cohen’s $d$, calculated from the posterior distributions between visit 1 and 2. ULR modeling indicated excellent prediction of response. 

Concerning the analysis of the DRO with the BTM, median parameter estimates were compared to the ground truth to assess the accuracy, otherwise not available with measured data. It was found that the Bayesian estimates generally have a very strong similarity with the ground truth, validating the accuracy of our BTM. The recovered parameters also have complementary regions of high and low percentage errors compared to the established NLLS fitting routine. RMS errors were lower for both implementations in the present work compared to similar DRO analysis by \citet{smith_dcemri.jl:_2015} and \citet{ortuno_dceurlab:_2013}. Albeit, the results are in good comparison. Caution is still required for voxels with low $v_e$. Concentration curves with these parameter combinations have very limited intensity changes which practically vanish in the added background noise.

The variance of estimates inherent in the Bayesian posterior distribution was compared to a bootstrapping error analysis, performed likewise to the work of \citet{kershaw_precision_2006}. It was demonstrated that the uncertainty maps of the BTM resemble those calculated with the bootstrap analysis, validating the precision of parameters recovered with the BTM. To the best of our knowledge, only \citet{schmid_bayesian_2006} implemented a Bayesian PK model with the objective to make use of the posterior probability distribution. They used it to state the probability of a pixel to be greater than a certain threshold for tumor masking. Their approach was applied to patient data, whereas in the present work, accuracy and precision of estimates were validated with a digital phantom first.

Furthermore, we applied the BTM to the breast cancer DCE-MRI data, performed PPCs and investigated the posterior distributions. For a PPC, the observed data was compared to the posterior predictive distribution, illustrated as percentile intervals of highest density. A good fit to the data results in a posterior distribution which reflects the noise level in the data; low noise corresponds to a narrow posterior and vice versa. However, a bad fit to the data results in a broad posterior distribution despite a low noise level. This indicates a systematic modeling error which influences the information we gained about uncertainty. More complex PK models which incorporate additional assumptions about CA transport, e.g. the extended Tofts model, could be able to produce a better fit to certain data. Hence, assessing posterior distributions requires to check the corresponding data and fit before drawing any conclusions from it. While feasible for ROI-based analysis with only a handful of concentration curves, visual assessment is not possible in a pixelwise analysis. An automated Bayesian model selection step as proposed in the work of \citet{duan_are_2017} could be an effective means to reduce systematic modeling error but is beyond the scope of this study.

In order to assess therapy response for the patients in the breast cancer DCE-MRI dataset, \citet{huang_variations_2014} showed in their original work that using visit 2 $K^\text{trans}$ or the percentage change of $K^\text{trans}$ between visits as metrics yields good to excellent results. However, the uncertainty in estimating PK parameters with tracer-kinetic models is not accounted for. For this purpose, we calculated Cohen's $d$ as a means of quantitative change in parameter estimates which depends on the precision of estimates. Using Cohen's $d$ metric, the assessment of response was found to be excellent by means of an ULR analysis. Considering the findings of the PPCs, including a model selection step as explained above could decrease the influence of systematic modeling errors on posterior distributions and hence Cohen's $d$ values which may further improve assessment of therapy response.

Limitations of the present work include large computational time when fitting the BTM to the DRO-data. On the one hand, the MCMC sampling is time and memory consuming but necessary to avoid divergences. On the other hand, it yields a full posterior probability distribution with information about the uncertainty, and obtaining the same information with a bootstrap analysis of a NLLS fit requires even more computation time. Furthermore, the simulated DRO curves have a much higher time-resolution compared to measured data. Evaluating real DCE-MRI data increases the speed of the analysis greatly. Moreover, the influence of the chosen prior distributions on the results was not assessed in the present study.

In conclusion, we evaluated a BTM with a DRO, assessed accuracy and precision against the standard NLLS approach and showed how posterior distributions are used to assess therapy response. We demonstrated that Bayesian modeling provides an elegant means to assess posterior probability distributions, which  are in good agreement with established approaches.


\section{Acknowledgments}
Funding: This work was supported by the research training group GRK 2274 of the DFG, Deutsche Forschungsgemeinschaft.


\appendix
\counterwithin{figure}{section}
\section{Prior Predictive Check}\label{app:prior_p_c}
\setcounter{figure}{0}
\begin{figure*}[ht!]
	\includegraphics[width=\linewidth]{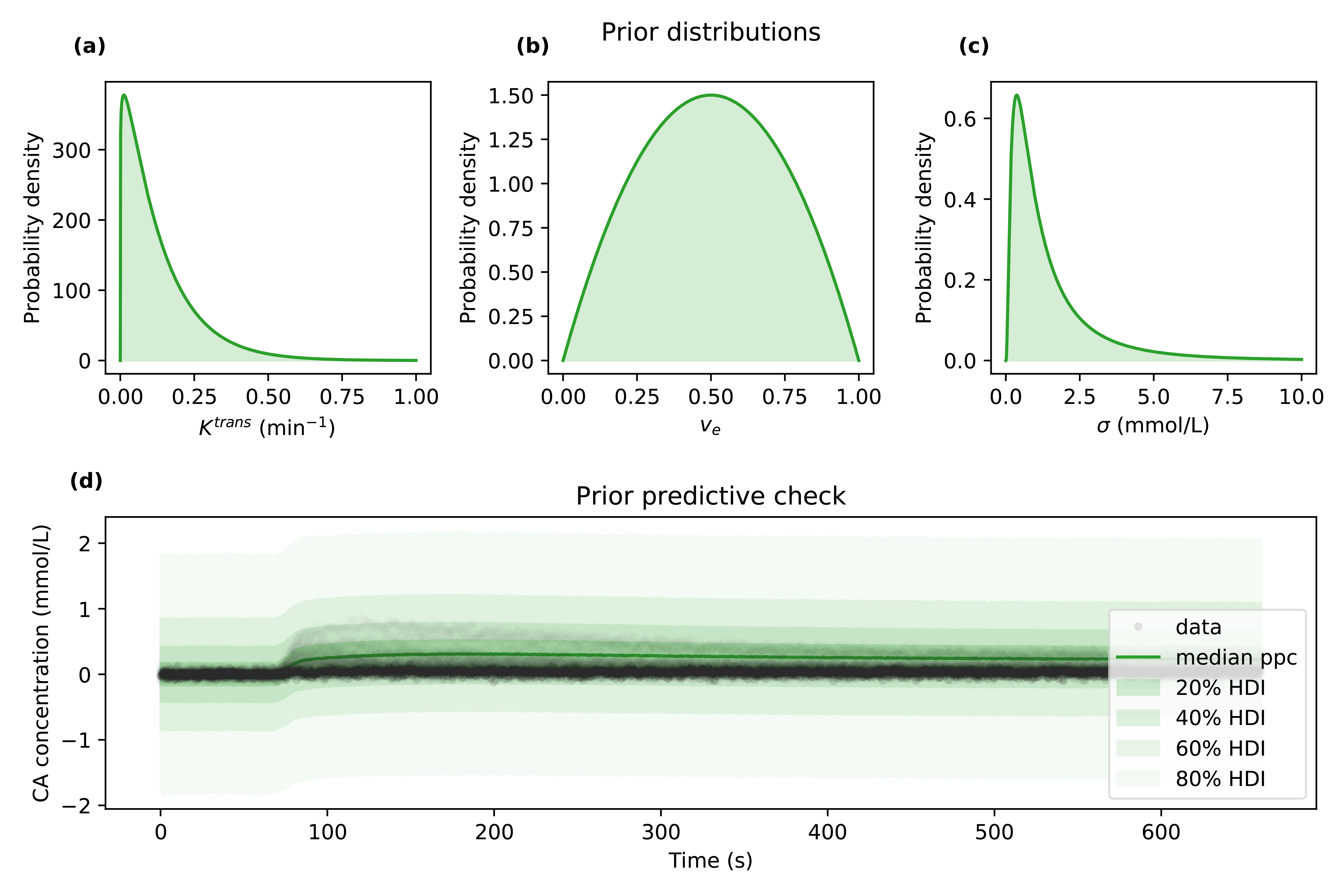}
	\caption{Prior probability density functions for $K^\text{trans}$ \textbf{(a)}, $v_e$ \textbf{(b)}, $\sigma$ \textbf{(c)} and the subsequent prior predictive check \textbf{(d)}. Simulated concentration-time curves (dots) for each parameter combination of the QIBA phantom are overlaid on the prior predictive distribution; illustrated by median (line) and the 20\% - 80\% highest density intervals (HDI).}
	\label{fig:priorpc}
\end{figure*}
To assess if the choice of prior distributions for the model parameters covers a reasonable range of concentration-time curves, it is useful to perform a \textit{prior predictive check}. For this purpose, we generated 100,000 MCMC samples from the prior predictive distribution,
\begin{equation}
P(\hat{y})=\int P(\hat{y}\mid \theta)\,P(\theta) \mathrm{d}\theta,
\end{equation}
only considering the prior distributions without any actual data. This quantifies the range of possible observations $\hat{y}$, predicted by our model. In a prior predictive check, the predicted data is compared to real observations and the extent of extreme observations indicates the level of disagreement between domain expertise and model assumptions. Fig. \ref{fig:priorpc} shows the probability density functions of the chosen priors \textbf{(a-c)} and the prior predictive check \textbf{(d)}. The black dots are actual observed data from the QIBA phantom, one curve for each parameter combination of $K^\text{trans}$ and $v_e$, to assess the scope of possible phantom curves. The increasingly lighter green bands represent the 20\%, 40\%, 60\% and 80\% highest density intervals between the corresponding percentiles of the prior predictive distribution; the green line is the median thereof. We find that the model predicts observations that are more extreme than the phantom data but not too extreme to be unrealistic given the assumed observational error. Hence, we conclude that the chosen prior distributions are reasonable.


\newpage
\bibliographystyle{model2-names.bst}
\bibliography{arxiv-bayes.bib}


\end{document}